\documentclass[aps, prx, reprint, superscriptaddress, longbibliography, floatfix]{revtex4-2}
\usepackage{graphicx}
\usepackage{dcolumn}
\usepackage{bm}
\usepackage{physics}
\usepackage{soul}
\DeclareMathAlphabet{\pazocal}{OMS}{zplm}{m}{n}
\DeclareMathOperator{\sgn}{sgn}
\usepackage{amsmath}
\usepackage{amsfonts}
\usepackage{mathrsfs}
\usepackage{subfigure}
\usepackage{color}
\usepackage{xcolor}
\usepackage[colorlinks=true,linkcolor=blue,anchorcolor=red,citecolor=blue, urlcolor=blue]{hyperref}
\setstcolor{red}

\begin{document}
\preprint{APS/123-QED}
 
\title{Nonequilibrium DC Current Generation in a Driven Dissipative Haldane Model}

\author {Konrad Koenigsmann}
\email{kk2abc@virginia.edu}
\affiliation{Department of Physics, University of Virginia, Charlottesville, Virginia 22904, USA}

\author{Sankha Subhra Bakshi}
\affiliation{Department of Physics, University of Virginia, Charlottesville, Virginia 22904, USA}

\author {Peter Schauss}
\affiliation{Department of Physics, University of Virginia, Charlottesville, Virginia 22904, USA}
\affiliation{Institut f\"ur Quantenphysik, Universit\"at Hamburg, Luruper Chaussee 149, 22761 Hamburg, Germany}

\author {Gia-Wei Chern}
\email{gc6u@virginia.edu}
\affiliation{Department of Physics, University of Virginia, Charlottesville, Virginia 22904, USA} 

\begin{abstract}
The interplay of topology with nonequilibrium driving and dissipation in open quantum systems has recently attracted significant interest in condensed matter physics. In this work, we investigate a driven, dissipative Haldane model using large-scale numerical simulations of Lindblad dynamics. We show that the system evolves into a time-periodic quasi-steady state when subjected to driving and dissipation, with 
nonequilibrium band occupations. 
To characterize this regime, we introduce an occupation-weighted Chern number as a nonquantized diagnostic constructed from the steady-state occupations and the geometry of the unperturbed bands. We further analyze charge transport in the presence of simultaneous driving and damping and demonstrate that a finite DC bulk current emerges when inversion symmetry is broken by a staggered sublattice potential. The magnitude and direction of this current are controlled by the driving amplitude, revealing a tunable nonequilibrium transport response 
enabled by inversion-symmetry breaking.
\end{abstract}

\date{\today}
\maketitle

\section{Introduction}
\label{sec:intro}

The dynamics of open quantum systems has emerged as a central topic of study in modern condensed matter physics, both to answer fundamental questions and drive technological progress. In contrast to isolated systems, open systems exchange energy and information with their environment, giving rise to nonequilibrium steady states that cannot be described using equilibrium statistical mechanics. Recent theoretical efforts have focused on understanding such steady states in interacting and noninteracting systems \cite{Prosen2010,Prosen2011,McDonald2023,Guimaraes2016,Santos2016}, the controlled preparation of quantum states and phases via engineered dissipation \cite{Diehl2008,Diehl2011,Kaczmarczyk2016}, and the emergence of genuinely dissipative many-body phenomena \cite{Wang2023,Paszko2024}. Parallel advances in experimental platforms have enabled realizations of open quantum simulators with tunable dissipation and driving \cite{Kaczmarczyk2016,Lee2019}, highlighting the relevance of these questions for quantum information science and nonequilibrium quantum technologies \cite{Reiter2017,Mueller2012,Aydogan2025}.

A particularly active direction concerns the fate of topological properties under nonequilibrium conditions, especially in the presence of periodic driving or coupling to an environment \cite{Mitra2024,Wang2023,Paszko2024,Yang2025}. While topological phases are traditionally defined using ground-state properties of closed systems, it is now well understood that driving and dissipation can fundamentally modify, degrade, or even stabilize topological features. The Haldane model \cite{Haldane1988}, which realizes a quantum Hall phase without Landau levels, provides a minimal and paradigmatic setting in which to explore these issues. Previous studies have examined the effects of either driving or dissipation on the Haldane model separately \cite{Yang2025,Wang2023,Mitra2024} as well as the behavior of simpler non-topological models under combined nonequilibrium protocols \cite{Santos2016,Prosen2011,Guimaraes2016}. 

Here, we investigate bulk DC transport in the Haldane model simultaneously subjected to coherent driving and dissipative relaxation modeled using the Lindblad formalism. 
The external drive is incorporated through a Peierls substitution, while dissipation is modeled using a relaxation-time approximation of a Lindblad master equation. Starting from an initial topological ground state, we show that the combined action of driving and damping evolves the system to a time-periodic quasi-steady state with nonequilibrium band occupations. 
To characterize this regime, we introduce an occupation-weighted Chern number that captures residual topological structure encoded in the nonequilibrium band populations. 

Beyond topology, we examine charge transport in the quasi-steady state and show that a finite DC bulk current is generated once inversion symmetry is broken by a staggered sublattice potential. This current appears in the absence of any static bias and is tunable via the driving amplitude, underscoring the crucial roles of symmetry breaking and dissipation in nonequilibrium transport. We further provide a physical interpretation of this effect in terms of intra- and inter-valley asymmetries in the nonequilibrium populations that fail to cancel and thereby produce a net current.

\begin{figure*}[t]
  \includegraphics[width=\textwidth]{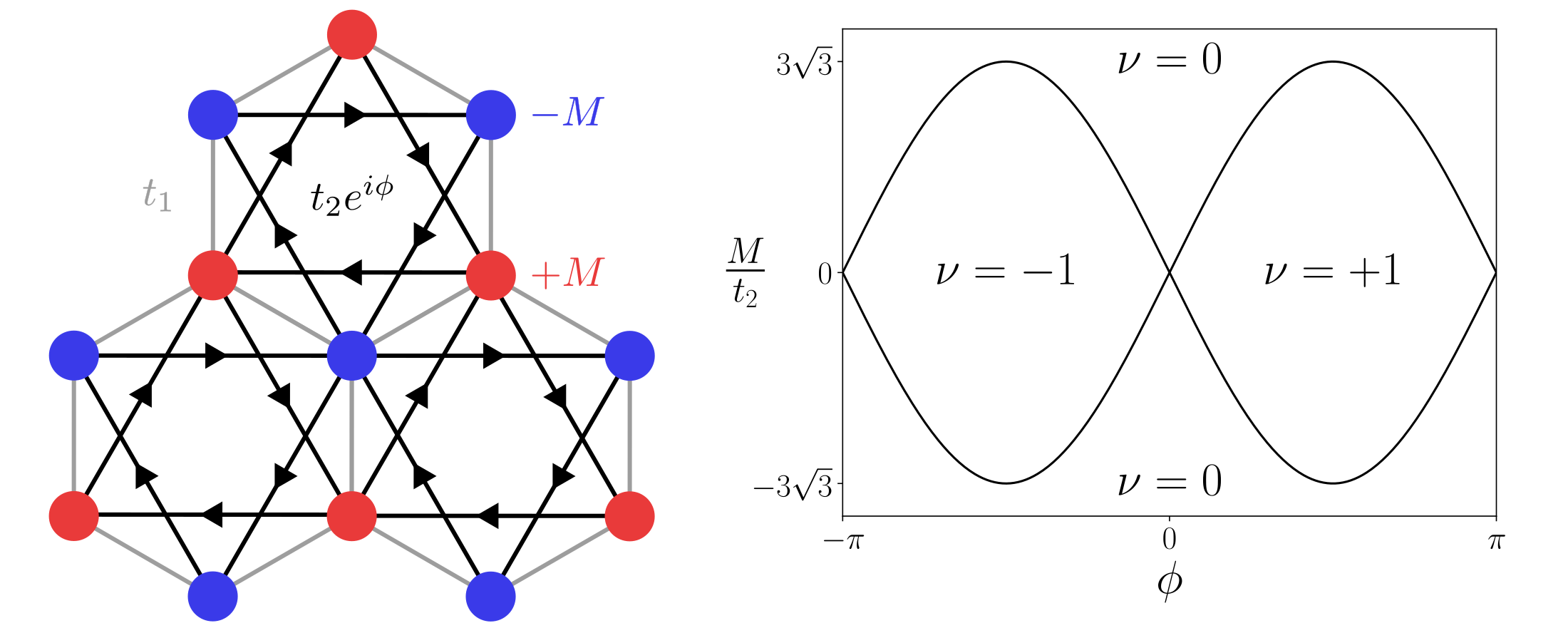}
  \caption{(a) The honeycomb lattice of the Haldane model with all salient features shown, including the nearest- and next-nearest-neighbor hoppings and the staggered sublattice potential. The two sublattices A and B are shown using red and blue sites, respectively. (b) Phase diagram of the Chern number of the Haldane model in the $\phi$-$M/t_2$ plane. $\nu=\pm1$ corresponds to the topological phase, and $\nu=0$ corresponds to the trivial phase.}
  \label{fig:1}
\end{figure*}
The remainder of this paper is organized as follows: In Sec. \hyperref[sec:sim]{II}, we introduce the Haldane model, the Peierls substitution, and the Lindblad master equation, along with the numerical methods used in our simulations. In Sec. \hyperref[sec:chern]{III}, we define the occupation-weighted Chern number and use it to characterize the time-periodic quasi-steady state. In Sec. \hyperref[sec:currents]{IV}, we present the calculation of DC currents in the quasi-steady state. Finally, in Sec. \hyperref[sec:discussion]{V}, we provide qualitative arguments explaining the emergence of a nontrivial DC bulk current in the quasi-steady state.

\section{Numerical Simulation of the Driven Damped Haldane Model}
\label{sec:sim}

\paragraph*{Driven damped Haldane model.} We begin by introducing the Haldane model in real space on the honeycomb lattice \cite{Haldane1988}. The Hamiltonian reads
\begin{equation}
\begin{split}
    \hat{H} = &-t_1\sum_{\langle i, j\rangle}\left(\hat{c_i}^\dagger\hat{c_j}+ \text{h.c.}\right)
    +t_2\sum_{\langle\langle i,j\rangle\rangle}\left(e^{i\phi_{ij}}\hat{c_i}^\dagger\hat{c_j}+\text{h.c.}\right)\\
    &+M\left(\sum_{i\in A}\hat{n_i} - \sum_{i\in B}\hat{n_i}\right),
\end{split}
\end{equation}
where $\hat{c}_i$ ($\hat{c}_i^\dagger$) annihilates (creates) a fermion at lattice site $i$, and $\hat{n}_i=\hat{c}_i^\dagger\hat{c}_i$ is the corresponding number operator. The first term describes nearest-neighbor hopping with amplitude $t_1$, while the second term represents next-nearest-neighbor hopping with amplitude $t_2$ and a complex phase $\phi_{ij}$, whose sign depends on the hopping direction, thereby breaking time-reversal symmetry. The third term corresponds to a staggered sublattice potential of strength $M$, which breaks inversion symmetry when $M\neq0$. The geometry and conventions are shown in Fig. \hyperref[fig:1]{1a}.

Since it is translationally invariant, the Haldane Hamiltonian can be Fourier transformed to momentum space, where it becomes block diagonal in crystal momentum $\mathbf{k}$. Each momentum sector corresponds to an independent two-level system in the sublattice $(A,B)$ basis,
\begin{equation}
H(\mathbf{k})
=
\mathbf{d}(\mathbf{k})\cdot\boldsymbol{\sigma}
=
\begin{pmatrix}
d_z(\mathbf{k}) & d_x(\mathbf{k})-i\,d_y(\mathbf{k}) \\
d_x(\mathbf{k})+i\,d_y(\mathbf{k}) & -d_z(\mathbf{k})
\end{pmatrix}.
\label{eq:Haldane}
\end{equation}
Here, the in-plane pseudospin components arise from nearest-neighbor hopping,
\begin{equation}
\begin{aligned}
d_x(\mathbf{k}) &= -t_1 \sum_{i=1}^{3} \cos(\mathbf{k}\cdot\boldsymbol{\delta}_i), \\
d_y(\mathbf{k}) &= t_1 \sum_{i=1}^{3} \sin(\mathbf{k}\cdot\boldsymbol{\delta}_i),
\end{aligned}
\end{equation}
\\
where ${\boldsymbol{\delta}_i}$ is the set of nearest-neighbor displacement vectors. The out-of-plane pseudospin component is generated by the staggered potential and complex next-nearest-neighbor hopping,
\begin{equation}
d_z(\mathbf{k})
=
M
+
2t_2\sin\phi
\sum_{i=1}^{3}
\sin(\mathbf{k}\cdot\mathbf{b}_i),
\end{equation}
where ${\mathbf{b}_i}$ is the set of next-nearest-neighbor displacement vectors. Here and below, we omit the scalar term
$d_0(\mathbf{k})I$, where
$d_0(\mathbf{k})=2t_2\cos\phi\sum_{i=1}^{3}
\cos(\mathbf{k}\cdot\mathbf{b}_i)$,
because it drops out of the density-matrix commutator.
The resulting band energies are
$E_{\pm}(\mathbf{k})
=d_0(\mathbf{k})\pm|\mathbf{d}(\mathbf{k})|$.

\begin{figure*}[!t]
  \includegraphics[width=\textwidth]{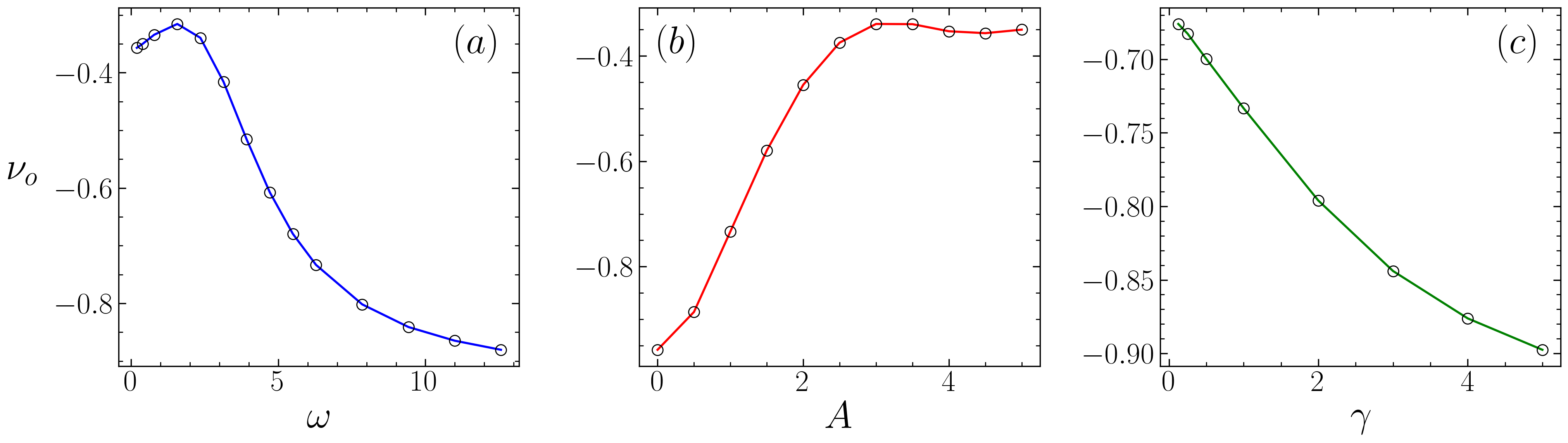}
  \caption{The occupation-weighted Chern number $\nu_o$ plotted as functions of (a) Driving frequency $\omega$; (b) Driving strength $A$; (c) Damping strength $\gamma$. For these results, $\omega=2\pi$, $A=1.0$, $\gamma=1.0$ unless shown otherwise in the plots.}
  \label{fig:occChern}
\end{figure*}

Near the Dirac points $\mathbf{K}$ and $\mathbf{K}'$, the Bloch Hamiltonian \eqref{eq:Haldane} reduces to the massive Dirac form
\begin{equation}
H_{\tau}(\mathbf{q}) = v_F(\tau q_x\sigma_x+q_y\sigma_y)+m_\tau\sigma_z,
\label{eq:Dirac}
\end{equation}
where $v_F$ is the Fermi velocity, $\tau=\pm 1$ is the valley index corresponding to $\mathbf{K}$ and $\mathbf{K}'$, respectively, and $\mathbf{q}=\mathbf{k}-\mathbf{K}_{\tau}$ is the momentum measured relative to the Dirac point $\mathbf{K}_{\tau}$. The Hamiltonian has valley-dependent masses $m_{\pm}=M\mp3\sqrt{3}t_2\sin\phi$. The equilibrium ground-state topology of the Haldane model is controlled by the competition between the staggered potential and the complex next-nearest-neighbor hopping. The system realizes a topological phase with nonzero Chern number $\nu$ when
\begin{equation}
    \left|M/t_2\right| < 3\sqrt{3}\left|\sin\phi\right|.
    \label{eq:Chernbound}
\end{equation}
More explicitly, the Chern number is given by \cite{Haldane1988}
\begin{equation}
\nu = \frac{1}{2}\left[\sgn(M+3\sqrt{3}t_2\sin\phi)-\sgn(M-3\sqrt{3}t_2\sin\phi)\right],
\end{equation}
as summarized in the phase diagram shown in Fig. \hyperref[fig:1]{1b}.

To model coupling to an environment, we describe the system dynamics using a Lindblad master equation \cite{Guimaraes2016, Manzano2020},
\begin{equation}
    \dv{\hat{\rho}}{t}=-i[\hat{H},\hat{\rho}] + \mathcal{D}(\hat{\rho}),
\end{equation}
where $\hat{\rho}$ is the many-body density matrix. The first term governs unitary evolution, while the dissipator $\mathcal{D}(\hat{\rho})$ encodes the effects of a thermal bath.

In this work, we adopt the relaxation-time approximation, in which dissipation drives the system toward a static equilibrium density matrix $\hat{\rho}^{eq}$,
\begin{equation}
    \mathcal{D}(\hat{\rho}) = 2\gamma\left[\hat{\rho}^{eq} - \hat{\rho}\right].
\end{equation}
Here $\gamma=1/2\tau$ sets the damping strength, with $\tau$ denoting the relaxation time, and $\hat{\rho}^{eq}$ is constructed from the outer products of the eigenstates of the static Hamiltonian. When expressed in momentum space, this approximation acts independently on each $\mathbf{k}$ sector, yielding a closed equation of motion for $\rho(\mathbf{k},t)$.

\paragraph*{Peierls substitution.} The system is driven out of equilibrium by an external AC electric field introduced via the Peierls substitution \cite{Peierls1933, Hofstadter1976}. In real space, the hopping amplitudes acquire a time-dependent phase,
\begin{equation}
    t_{ab}\to t_{ab}\exp\left(i\int_{r_a}^{r_b}\mathbf{A}(t)\cdot\dd \boldsymbol{\ell}\right),
\end{equation}
where $\mathbf{A}(t)$ is the vector potential related to the electric field through $\mathbf{E}(t)=-\partial_t\mathbf{A}(t)$. We take
\begin{equation}
\mathbf{A}(t)=A\cos(\omega t)\,\hat{\mathbf e},
\end{equation}
with amplitude $A$, frequency $\omega$, and polarization direction $\hat{\mathbf e}$. Since $\mathbf{A}(t)$ is spatially uniform, the line integral reduces to $\mathbf{A}(t)\cdot\Delta\mathbf{r}$. 

In momentum space, this prescription corresponds to a minimal substitution,
\begin{equation}
\mathbf{k}\rightarrow\mathbf{k}-\mathbf{A}(t),
\end{equation}
so that each momentum block evolves under a time-dependent Bloch Hamiltonian
\begin{equation}
H(\mathbf{k},t)=\mathbf{d}(\mathbf{k}-\mathbf{A}(t))\cdot\boldsymbol{\sigma}.
\end{equation}
As a result, the full driven-dissipative problem reduces to a collection of independent, driven two-level density matrices $\rho(\mathbf{k},t)$ evolving in parallel in momentum space.

\paragraph*{Units and normalization.}
Throughout this work, we set $e=\hbar=1$ and $t_1\equiv 1$. Therefore, the Hamiltonian parameters $t_2, M$ and energies $\epsilon$ are expressed in terms of $t_1$, and the damping strength $\gamma$ and frequency $\omega$ are expressed in terms of $\hbar/t_1$. The vector potential strength $A$ is given in terms of $ea/\hbar$, where $a$ is the nearest-neighbor bond length, which we have set to 1 here.

\paragraph*{Numerical simulation.} We solve the Lindblad equation numerically using a fourth-order Runge--Kutta scheme, both in real space and in momentum space, with a timestep of 0.01 $\hbar/t_1$. The system is initialized with a fully occupied valence band and an empty conduction band. Unless otherwise stated, the Hamiltonian parameters were chosen as $t_2=0.2$, $\phi=-\pi/4$, and $M=0.1$. Real-space simulations were performed on a $13\times13$ lattice, while momentum-space calculations used grids up to $100\times100$. The driving field was polarized in the positive $x$ direction.

\begin{figure*}[!t]
  \includegraphics[width=\textwidth]{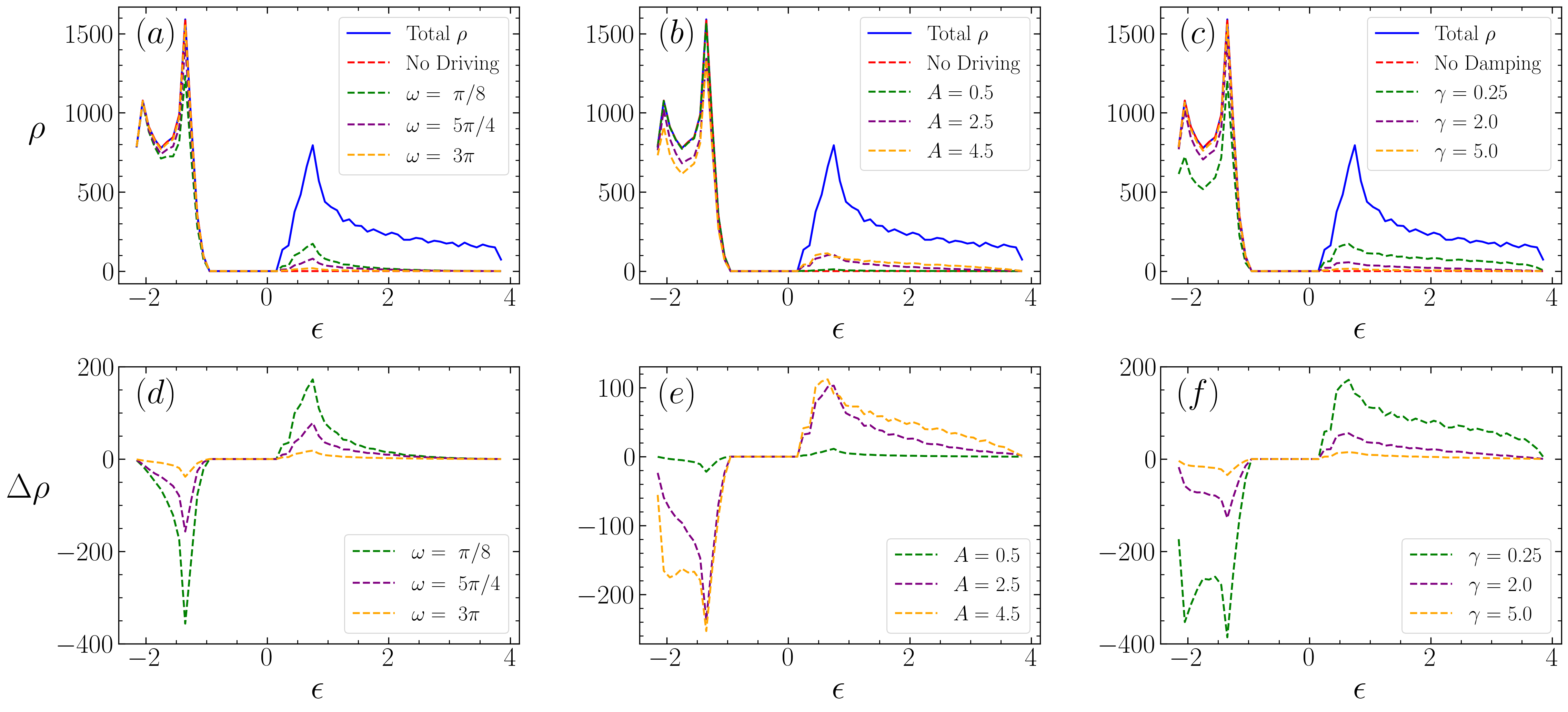}
  \caption{The occupations of the driven dissipative Haldane model change relative to half-filling when nonzero driving and damping are applied. Here, $\epsilon$ denotes energy, and $\rho$ denotes the density of states. The top row displays the absolute changes in occupations as functions of (a) Driving frequency $\omega$; (b) Driving strength $A$; (c) Damping strength $\gamma$. Half-filling, depicted in these plots with a red dashed line, shows that the valence band is filled and the conduction band is empty, while both bands are partially filled when the driving field is turned on. The bottom row displays the relative changes in occupations relative to half-filling as functions of (d) Driving frequency $\omega$; (e) Driving strength $A$; (f) Damping strength $\gamma$. For these results, $\omega=2\pi$, $A=1.0$, $\gamma=1.0$ unless shown otherwise in the plots.}
  \label{fig:cherndos}
\end{figure*}

Due to the periodic drive, the system does not relax to a static steady state but instead evolves to a time-periodic quasi-steady state with period $T=2\pi/\omega$ \cite{Gao2022}. All observables reported in the following sections are therefore obtained by averaging the density matrix over one full driving period in this long-time regime.

\section{Occupation-Weighted Chern Number}
\label{sec:chern}

We first consider the effects of the driving and damping forces on the topological phase of the Haldane model. By preparing the system at half-filling with the Hamiltonian parameters listed in the previous section, the system is in the topological phase with a nonzero Chern number. 
The Chern number can be computed in momentum space using the TKNN formula \cite{Thouless1982}:
\begin{equation}
    \nu(P)=\frac{1}{2\pi i}\int_{\text{BZ}}\Tr\left(P\left(\pdv{P}{k_x}\pdv{P}{k_y}-\pdv{P}{k_y}\pdv{P}{k_x}\right)\right)\dd k_x \dd k_y,
    \label{eq:TKNN}
\end{equation}
where $P$ is the spectral projector and is related to the density matrix as follows: $\rho_{jk}=\bra{\psi}\hat{c_j}^\dagger\hat{c_k}\ket{\psi}=P_{kj}$. The trace is taken over the band indices. However, this formula can only be used to compute the Chern number from the density matrix if $\hat{\rho}^2=\hat{\rho}$ so that it satisfies the definition of a projector.

\begin{figure*}
  \includegraphics[width=\textwidth]{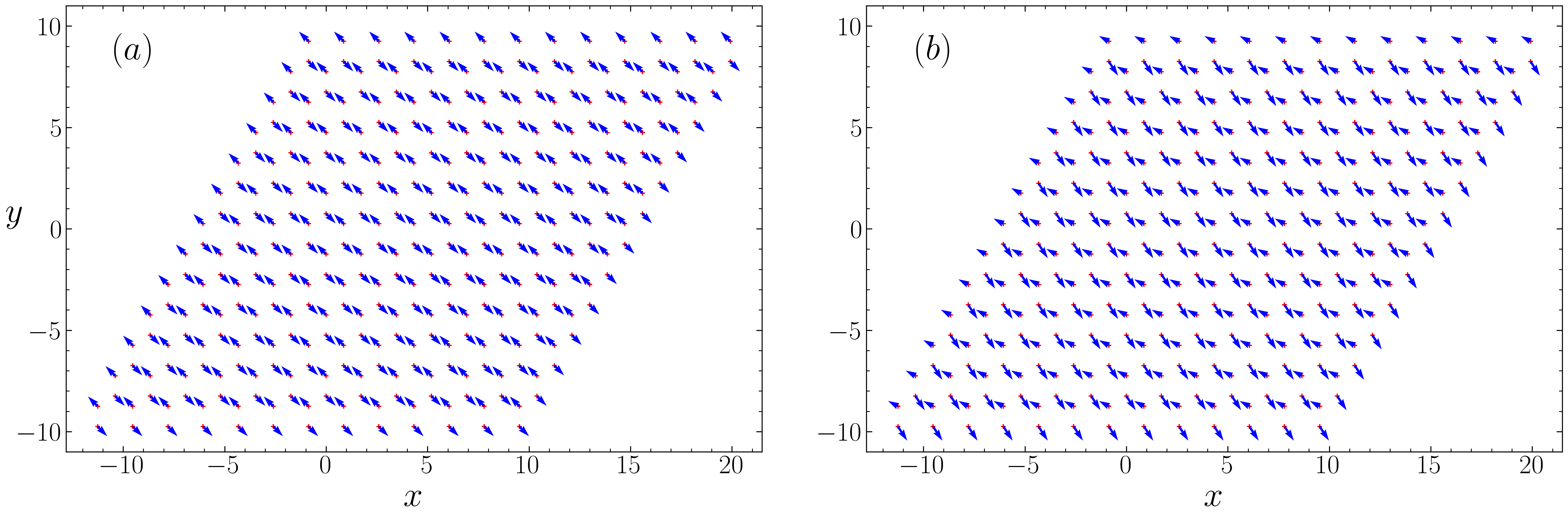}
  \caption{The net current per lattice site plotted for the cases of (a) $M=0$; (b) $M=0.1$. The red crosses denote the lattice sites, and the blue arrows denote the direction and relative magnitude of the current at each lattice site. When $M=0$, the unit-cell averaged net current is $7.12 \times10^{-9}$, and when $M=0.1$, the unit-cell averaged net current is $1.32\times10^{-2}$.}
  \label{fig:netcurrentlattice}
\end{figure*}

Upon reaching the quasi-steady state, the time-averaged density matrix is no longer a projector, and hence the TKNN formula cannot be used to compute the Chern number of the quasi-steady state. On the one hand, the fact that this density matrix is no longer a projector indicates that the system is no longer in a topological phase and that the system cannot be characterized by a quantized nonzero Chern number. On the other hand, such a quasi-steady state is not generally equivalent to the trivial equilibrium ground state with Chern number zero \cite{Dehghani2014, Iadecola2015}.

In order to characterize this quasi-steady state, we introduce the occupation-weighted Chern number $\nu_o$. First, we define the occupation-weighted density matrix:
\begin{equation}
    \hat{\rho}_o=\sum_k\sum_{n}f_n(k)\rho_n(k),
\end{equation}
where $n$ is the band index, $f_n(k)$ is the occupation of band $n$ at momentum $\mathbf{k}$, as calculated from the time-averaged density matrix, and $\rho_n(\mathbf{k})=\ket{u_n(\mathbf{k})}\bra{u_n(\mathbf{k})}$. Here $\{\ket{u_n(\mathbf{k})}\}$ are the Bloch eigenstates of the unperturbed Haldane Hamiltonian. The occupations are normalized such that $\sum_n f_n(k) = 1$ in the case of half-filling. Then $\nu_o$ is calculated using \eqref{eq:TKNN} with $\hat{\rho}_o^\dagger$, which is not a true projector, in place of the spectral projector $P$.

In this work, $\nu_o$ is numerically calculated over a discretized Brillouin zone, with the derivatives of $\hat{\rho}_o^\dagger$ at each point evaluated using finite differences. We note that $\hat{\rho}_o^\dagger$ is invariant under a gauge transformation $\ket{u_n(\mathbf{k})}\to e^{i\theta_n(\mathbf{k})}\ket{u_n(\mathbf{k})}$. We further note that at the system sizes used in our simulations, the Chern number for the system at half-filling calculated using this method is not quantized, but we have verified that the Chern number converges to a quantized value as the discretization is made finer.

The occupation-weighted Chern numbers as a function of the driving frequency $\omega$, the vector potential strength $A$, and damping strength $\gamma$ are shown in Fig. \ref{fig:occChern}.  Over most of the parameter space, $\nu_o$ grows monotonically with $A$ and decreases monotonically with $\omega$ and $\gamma$. 
Specifically, $\nu_o$ approaches a constant value in the limits $\omega\to\infty,A\to0,$ or $\gamma\to\infty$. As $\omega\to\infty$, the period approaches zero and the high-frequency oscillations of the drive can be averaged over one period. 
To leading order in the high-frequency expansion, this averaging yields an effective time-independent Hamiltonian. 
This leads to a static renormalization of the nearest- and next-nearest-neighbor hoppings, meaning that the system is in the topological phase in this limit if \eqref{eq:Chernbound} still holds. As $A\to0$, the driving strength becomes weaker relative to the damping strength. Therefore, deviations of the time-averaged density matrix in the quasi-steady state from the static equilibrium density matrix will be smaller, and the system approaches the topological phase in this limit. The effect is the same in the $\gamma\to\infty$ limit, except that in this case it stems from the fact that the damping strength becomes stronger relative to the driving strength rather than the driving strength becoming weaker.

To better understand how the occupation-weighted Chern number characterizes the quasi-steady state, the occupations as functions of $\omega, A$, and $\gamma$ can be examined directly (Fig. \ref{fig:cherndos}). Relative to half-filling (the red line in Figs. \hyperref[fig:cherndos]{3a-c}), which shows a completely filled valence band ($\epsilon<0$) and an empty conduction band ($\epsilon>0$), when the driving is turned on, some states in the valence band are excited to the conduction band, so that there are now two partially filled bands. This is seen more clearly in Figs. \hyperref[fig:cherndos]{3d-f}, which show the occupations relative to half-filling when the driving is turned on. The lower occupations at negative energies are offset exactly by the higher occupations at positive energies. The changes in occupation are monotonic as a function of the driving parameters, approaching zero change in the limits $\omega\to\infty, A\to0,$ and $\gamma\to\infty$, in agreement with Fig. \ref{fig:occChern}. 

This confirms that the occupation-weighted Chern number is a suitable quantity for characterizing the quasi-steady state. While this state is no longer topological and therefore does not have a quantized nonzero Chern number, it is clearly not in the trivial state with zero Chern number. The occupation-weighted Chern number reflects the fact that the state now has two partially occupied bands, rather than one filled and one empty band, an indication that the quasi-steady state retains some residual topology.

\section{DC Currents of the Quasi-Steady State}
\label{sec:currents}

The quasi-steady state can additionally be characterized by the response of physical observables, such as the charge current, to the driving and damping forces. For a single bond, the current operator is given by \cite{Mardanya2018}:
\begin{equation}
    \hat{\mathbf J}_{ij}
    =
    i\boldsymbol{\delta}_{ij}\left(t_{ij}\hat{c}_i^\dagger\hat{c}_j - \text{h.c.}\right)
    =
    -2\boldsymbol{\delta}_{ij}\Im\!\left(t_{ij}\hat{c}_i^\dagger\hat{c}_j\right),
\end{equation}

\begin{figure*}[!htb]
\includegraphics[width=\textwidth]{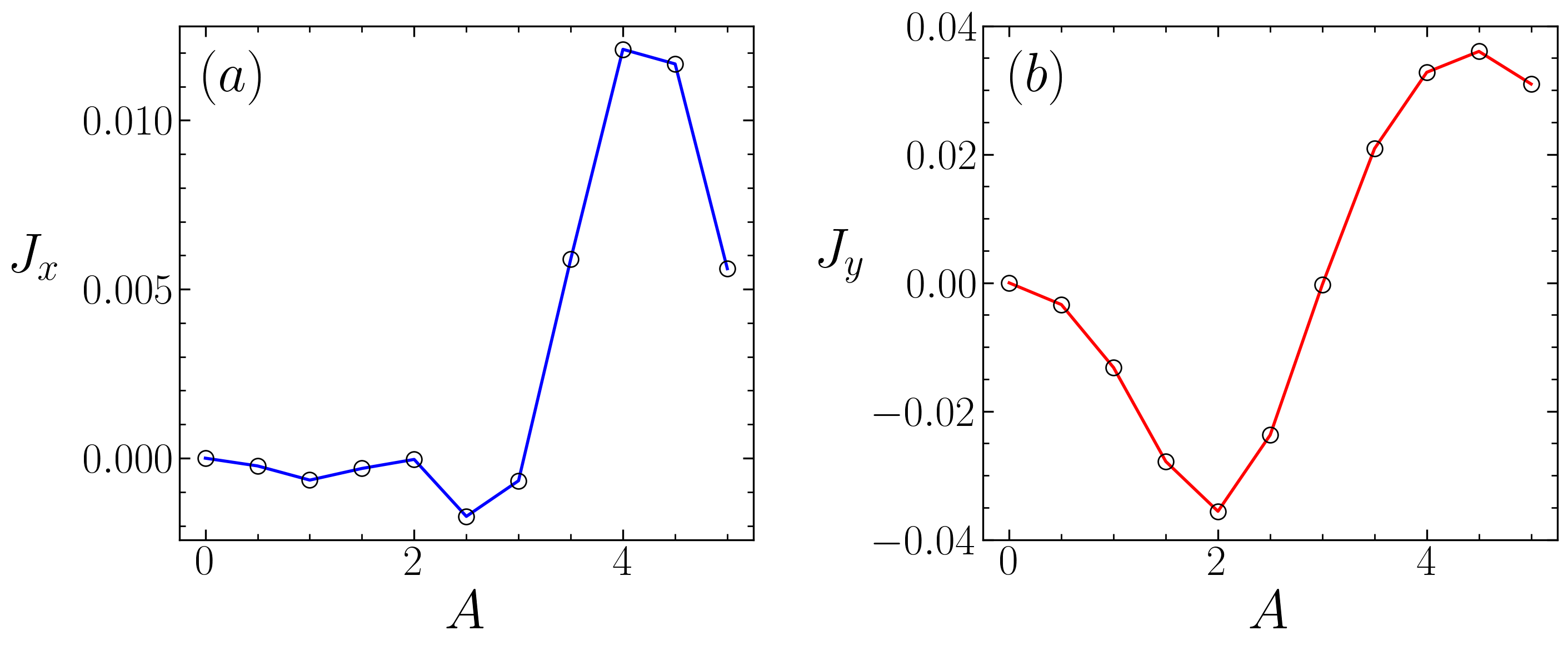}
  \caption{The (a) $x$- and (b) $y$-components of the unit-cell–averaged net current plotted as functions of the driving strength $A$. The direction of the current reverses near $A=3$.}
  \label{fig:netcurrentcomps}
\end{figure*}

\noindent where $\boldsymbol{\delta}_{ij}$ is the displacement vector connecting sites $i$ and $j$, and $t_{ij}$ is the hopping amplitude between them. Following the normalization defined in Sec. \ref{sec:sim}, the bond-current is given in units of $et_1a/\hbar$. The expectation value of this operator can be computed directly from the single-particle density matrix. 
Since both $\langle \hat{c}_i^\dagger\hat{c}_j \rangle$ and $t_{ij}$ are time-dependent in the presence of driving, the time-averaged current is obtained by evaluating $\langle \hat{\mathbf{J}}_{ij}(t)\rangle$ at each timestep after the system reaches the quasi-steady state and subsequently averaging over one full driving period.

The net current operator at site $i$ is defined as $\hat{\mathbf{J}}_i=\sum_j\hat{\mathbf{J}}_{ij}$, where the sum runs over all nearest-neighbor and next-nearest-neighbor bonds connected to site $i$. It is also useful to consider the net current averaged over a unit cell. Since the honeycomb lattice has a two-site basis, the unit-cell–averaged current is obtained by summing the average net currents on sublattices A and B. 
This unit-cell–averaged quantity is the physically relevant measure of bulk transport, as it properly takes into account lattice symmetries.

Fig. \ref{fig:netcurrentlattice} shows representative examples of the net current per site in the quasi-steady state when both driving and damping are present. In the static Haldane model (not shown), broken time-reversal symmetry leads to circulating next-nearest-neighbor bond currents that sum to zero at every lattice site. When driving and damping are turned on, in contrast, nonzero nearest-neighbor bond currents appear and the components of the next-nearest-neighbor bond currents are no longer all equal. As a result, the net current evaluated at individual lattice sites becomes nonzero. 

It has been previously established that a finite DC current can be generated in periodically driven systems when both time-reversal and inversion symmetries are broken \cite{Gao2022}. In the special case where only inversion symmetry is broken, a DC response has been observed in other systems \cite{Sotome2019} and is commonly referred to as the shift current \cite{Resta2024}. 
If one considers only the site-resolved net current in the present model, it may appear that a DC current is generated even when inversion symmetry is preserved (Fig. \hyperref[fig:netcurrentlattice]{4a}). 
This apparent contradiction is resolved by noting that symmetry constraints apply to the unit-cell–averaged current rather than to site-resolved currents.
Indeed, when inversion symmetry is preserved, the unit-cell–averaged current vanishes identically. Only when both time-reversal and inversion symmetries are broken (Fig. \hyperref[fig:netcurrentlattice]{4b}) does a finite unit-cell–averaged DC current emerge.

We further characterize this DC current by computing the unit-cell–averaged current as a function of the driving strength $A$, as shown in Fig. \ref{fig:netcurrentcomps}. The $x$- and $y$-components of the current are plotted separately in order to capture both changes in magnitude and direction. At all driving strengths, $|J_y|>|J_x|$. This is because the transverse velocity of Bloch electrons in weak electromagnetic fields contains an anomalous velocity term proportional to the Berry curvature \cite{Sundaram_1999}. Since a highly uneven Berry curvature distribution arises when both time-reversal and inversion symmetries are broken, the transverse velocity can be much larger than the longitudinal velocity, and thus the transverse current would be larger than the longitudinal current as well. 

In both $J_x$ and $J_y$, a reversal of the current direction is observed near $A=3$. To understand this behavior, we first consider the high-frequency limit of the driven system. As discussed in Sec. \ref{sec:chern}, in the limit $\omega\to\infty$ the driving renormalizes the hopping amplitudes. This can be written as $t_i^{\mathrm{eff}} \sim t_i J_0(\alpha)$,
where $J_0$ is the zeroth-order Bessel function and $\alpha = A\,\hat{e}\cdot\Delta\mathbf{r}$ \cite{Bukov2015}. Consequently, the effective hopping amplitudes change sign when $J_0(\alpha)$ crosses one of its zeros, leading to a reversal of the current direction.

\begin{figure*}[!htb]
  \includegraphics[width=\textwidth]{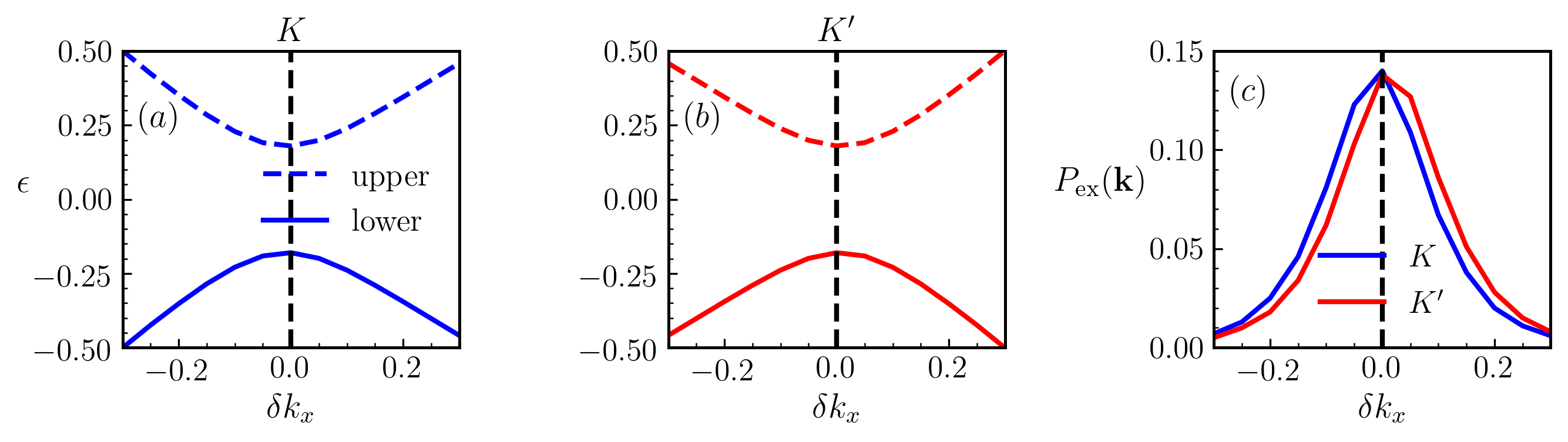}
  \includegraphics[width=\textwidth]{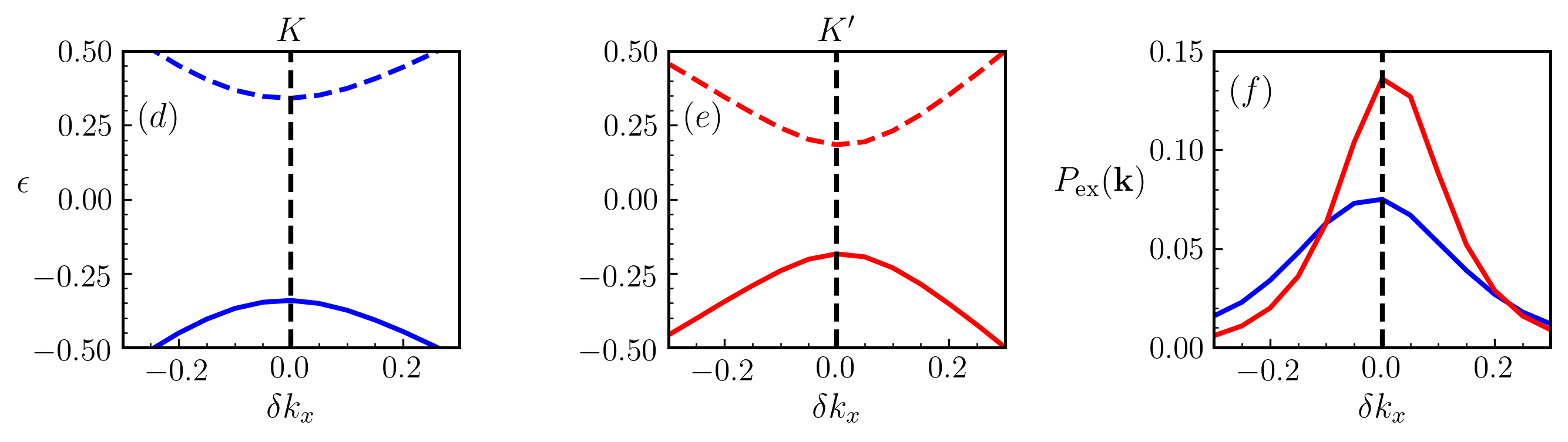}
  \caption{
Momentum-resolved band dispersion and excited-state population for the driven damped Haldane model for (a)–(c): $\phi=0$ ($\nu=0$); (d)–(f): $\phi=\pi/2$ ($\nu=1$). Here, $\epsilon$ denotes energy and $\delta k_x$ denotes the relative difference in the $x$-component of crystal momentum to $\mathbf{K}$ or $\mathbf{K}'$.
(a,b) Band dispersion near the two inequivalent valleys $K$ and $K'$, which possess identical gaps in the trivial phase.
(c) Corresponding excitation measure $P_{\rm ex}(k)$ plotted for both valleys. Although intra-valley population asymmetry develops, the excitation remains valley-symmetric overall, leading to valley currents that cancel each other and zero net bulk current.
(d,e) The dispersions at $K$ and $K'$ now exhibit inequivalent gaps in the topological phase.
(f) The excitation profiles at the two valleys are no longer related by valley symmetry, generating unequal valley currents that do not cancel, resulting in a finite bulk DC current.
For all results shown here we take $t_1 = 1$, $t_2 = 0.04$, $M=0.1$, $E_0=0.1$, $\omega =0.22$, and $\gamma=0.2$.}
\label{fig:phi}
\end{figure*}

Although the parameters used in Fig. \ref{fig:netcurrentcomps} do not strictly correspond to the high-frequency regime, the same physical mechanism remains operative provided higher-order terms in the Magnus expansion are small. This assumption can be tested numerically by comparing currents computed using $\overline{t_{ij}\langle\hat{c}_i^\dagger\hat{c}_j\rangle}$ with those obtained from $\overline{t}_{ij}\,\overline{\langle\hat{c}_i^\dagger\hat{c}_j\rangle}$, where the overline denotes time averaging. For the parameter regime considered here, the two procedures yield quantitatively similar results, indicating that higher-order frequency corrections are negligible and that the observed current reversal can be attributed primarily to the sign change of the effective hoppings induced by the drive.

\section{Qualitative interpretation of the bulk DC current}
\label{sec:discussion}
We present two qualitative arguments for understanding the origin of the finite DC current introduced in the previous section, first from a real-space perspective and then from a momentum-space perspective.

\paragraph*{Real space.} Within the driven dissipative Haldane model, the origin of the unit-cell-averaged DC current can be understood in real-space terms as follows: The staggered sublattice potential breaks inversion symmetry and renders the two sublattices energetically inequivalent. While this potential does not directly modify the hopping amplitudes, it leads to different local energetics and wavefunction weights on the two sublattices, which in turn produce inequivalent nonequilibrium charge responses under driving. As a result, when the system is subjected to a periodic electric field and relaxes to a quasi-steady state, the redistribution of charge between the two sublattices 
can contribute to a finite unit-cell-averaged DC current. 
The accompanying changes in electronic polarization suggest a possible connection to the shift-current mechanism previously discussed in the literature \cite{Gao2022,Fregoso2017,Dai2023}.

A complementary qualitative perspective is provided by the nonequilibrium populations in momentum space. Therefore, we now analyze the driven steady state in the vicinity of the two inequivalent valleys $K$ and $K'$, and 
discuss how asymmetries in the valley-resolved nonequilibrium populations 
can contribute to uncompensated valley currents and a finite bulk transport response.

\paragraph*{Nonequilibrium occupation dynamics.} For each crystal momentum we define the 
excitation measure
\begin{equation}
P_{\rm ex}(\mathbf{k})
=2f_+(\mathbf{k}),
\end{equation}
where $f_{\pm}(\mathbf{k})$ denote the occupations of the upper and lower bands, respectively. In equilibrium, $P_{\rm ex}=0$, while optical pumping (for example, the external AC electric field we use to drive the system out of equilibrium here) creates a finite 
excitation measure with a strongly momentum-dependent structure, as shown in Figs. \hyperref[fig:phi]{6a--c} for $\phi=0$ and in Figs. \hyperref[fig:phi]{6d--f} for $\phi=\pi/2$.

To illustrate how an asymmetric population distribution can contribute to transport, we consider the following estimate based on the group velocities of the unperturbed bands.
\begin{equation}
\mathbf{J}_{pop}
=
-e
\sum_{n=\pm}
\int_{\rm BZ}
\frac{d^2k}{(2\pi)^2}\,
f_n(\mathbf{k})\,\mathbf{v}_n(\mathbf{k}),
\label{eq:J_general}
\end{equation}
where
\begin{equation}
\mathbf{v}_n(\mathbf{k})
=
\nabla_{\mathbf{k}}E_n(\mathbf{k})
\end{equation}
is the semiclassical band velocity, $E_{\pm}(\mathbf{k})$ being the energy dispersion.
In terms of the valley-resolved momenta $\mathbf{q}$ (see \eqref{eq:Dirac}), we estimate the contribution from a symmetric valley region $\mathcal{V}_\tau$ centered on $K_\tau$:
\begin{equation}
\mathbf{J}_{pop,\tau}
=
-e
\sum_{n=\pm}
\int_{\mathcal{V}_\tau}
\frac{d^2q}{(2\pi)^2}
\,f_{n,\tau}(\mathbf{q})\,
\nabla_{\mathbf{q}}E_{n,\tau}(\mathbf{q}).
\end{equation}
Within the isotropic Dirac approximation, since $\nabla_{\mathbf{q}}E_{n,\tau}(\mathbf{q})\!\propto\!(\cos\varphi,\sin\varphi)$, $\varphi$ being the azimuthal polar angle,
only the \textit{odd} angular component of the nonequilibrium distribution,
$f(\mathbf{q})\neq f(-\mathbf{q})$,
contributes to transport. Thus, an intra-valley population asymmetry can contribute to
$\mathbf{J}_{pop,\tau}$.

For $\phi=0$, the Haldane model is topologically trivial and preserves an effective valley symmetry,
\begin{equation}
\begin{aligned}
E_{n,K}(\mathbf{q}) &= E_{n,K'}(-\mathbf{q}), \\
\nabla_{\mathbf{q}} E_{n,K}(\mathbf{q})
&= - \nabla_{\mathbf{q}} E_{n,K'}(-\mathbf{q}) .
\end{aligned}
\end{equation}
Here the gradients are evaluated at the indicated momenta.

Optical pumping indeed produces intra-valley asymmetry,
$f(\mathbf{q})\neq f(-\mathbf{q})$,
as visible in the momentum-dependent excitation profiles in Fig. \hyperref[fig:phi]{6c}, but
the corresponding population contributions cancel if $f_{n,K}(\mathbf{q})=f_{n,K'}(-\mathbf{q})$ over symmetry-related valley regions:
\begin{equation}
\mathbf{J}_{pop,K}
= -\mathbf{J}_{pop,K'}.
\end{equation}
This condition yields no net contribution from these valleys to the population-based current estimate.

For $\phi=\pi/2$ and the nonzero staggered potential considered here, time-reversal symmetry is broken and the Dirac gaps at the two valleys differ,
\begin{equation}
\Delta_K \neq \Delta_{K'}, \quad \Delta_\tau=2|m_\tau|.
\end{equation}
Because the excitation measure depends sensitively on the local gap and curvature, the photoexcited populations become valley-asymmetric:
\begin{equation}
f_{n,K}(\mathbf{q}) \neq f_{n,K'}(-\mathbf{q}).
\end{equation}
Such asymmetry can permit an uncompensated valley contribution to the population-based estimate,
\begin{equation}
\begin{aligned}
&\mathbf{J}_{pop}\approx\mathbf{J}_{pop,K}+\mathbf{J}_{pop,K'},
\end{aligned}
\end{equation}
where the approximation retains the contributions near the two valleys. This qualitative change in the excitation profiles is illustrated by comparing Fig. \hyperref[fig:phi]{6c} with Fig. \hyperref[fig:phi]{6f}: for $\phi=\pi/2$, the excitation profiles around $K$ and $K'$ are no longer symmetry-related,
suggesting a possible source of incomplete cancellation between valley contributions.

The full DC current reported in the previous section is evaluated directly from the time-dependent bond-current operator and density matrix. It includes interband coherences and the explicit time dependence of the current operator, which are not included in the population-based estimate above. The valley picture therefore provides a qualitative interpretation rather than a quantitative decomposition of the full current.

\section{Conclusion}
\label{sec:conc}
In this work, we have examined some properties of a driven dissipative Haldane model using large-scale numerical simulations. By characterizing the system using the occupation-weighted Chern number, we found that when initializing the system at half-filling in the topological phase of the Haldane model, the system relaxes to a quasi-steady state that is neither a topological phase nor a trivial phase due to shifts in occupations from a filled valence band and empty conduction band to partially filled valence and conduction bands. This occupation-weighted Chern number approaches a constant value in the high driving frequency, low driving strength, and high damping strength limits as expected, validating this quantity as suitable for characterizing the quasi-steady state and indicating that this quasi-steady state retains some residual topology.

We have also characterized the response of the charge current to the introduction of driving and damping forces in the Haldane model. When inversion symmetry is broken via a nontrivial staggered sublattice potential, a net DC current is generated in the system.
This can be understood as the result of a preferential flow of electric charge from one sublattice to the other due to inequivalent nonequilibrium charge responses under driving. 
The microscopic origin of this DC current can be understood as an asymmetry in the nonequilibrium occupations near the $K$ and $K'$ valleys. By examining the net DC current averaged over one unit cell, we confirm that a net DC current is only generated when inversion symmetry is broken. We also observe a reversal in the direction of this net DC current as the driving strength changes. This can be understood by approximating the effective tunneling strengths as being proportional to the zeroth Bessel function, which is valid in the regime we considered here, since the Bessel function changes sign as it passes through its zeroes. 

The findings presented here lay the groundwork for further examination of this driven dissipative Haldane model and characterization of its phase diagrams through simulations over a broader parameter space. One particularly intriguing question not answered by the present work is how the above results would change if an interface were introduced to this model. In the unperturbed case, a chiral edge current emerges at the interface, so one would expect that the introduction of damping and driving forces leads to an interplay between the net DC current generated by broken inversion symmetry and the interface edge current.

\begin{acknowledgments}
We thank Jeffrey Teo for insightful discussions on the theoretical aspects of the driven damped Haldane model. This work was supported by the Owens Family Foundation. K.K, S.S.B., and G.-W. C. acknowledge Research Computing at The University of Virginia for providing computational resources and technical support.
\end{acknowledgments}
\section*{Conflict of interest statement}
The authors have no conflicts to disclose.

\section*{Authorship contribution statement}
Konrad Koenigsmann: Writing---original draft, conceptualization, investigation, formal analysis, software. Sankha Subhra Bakshi: Writing---original draft, formal analysis. Peter Schauss: Writing---review and editing, supervision. Gia-Wei Chern: Writing---review and editing, conceptualization, supervision, funding acquisition.

\section*{Data availability statement}
The data that support the findings of this study are available from the corresponding author upon reasonable request.

\raggedbottom
\bibliography{ref}
\end{document}